# Illustrating an Effective Workflow for Accelerated Materials Discovery

Authors: Mrinalini Mulukutla[a] *, A. Nicole Person[a], Sven Voigt[b], Lindsey Kuettner[d], Branden Kappes[d], Danial Khatamsaz[a], Robert Robinson[a], Daniel Salas[a], Wenle Xu[a], Daniel Lewis[a], Hongkyu Eoh[a], Kailu Xiao[a], Haoren Wang[e], Jaskaran Singh Saini[e], Raj Mahat[b], Trevor Hastings[a], Matthew Skokan[a], Vahid Attari[a], Michael Elverud[a], James D. Paramore[a,g], Brady Butler[a,g], Kenneth Vecchio[f], Surya R. Kalidindi[b,c], Douglas Allaire[f], Ibrahim Karaman[a], Edwin L. Thomas[a], George Pharr[a], Ankit Srivastava[a], Raymundo Arróyave[a]

[a] Department of Materials Science and Engineering, Texas A&M University, College Station, TX 77843, USA

[b] School of Materials Science and Engineering, Georgia Institute of Technology, Atlanta, GA 30332-0245, United States

[c] George W. Woodruff School of Mechanical Engineering, Georgia Tech, Atlanta, GA 30332, USA

[d] Contextualize, LLC, Littleton, CO, USA

[e] Department of NanoEngineering, University of California, San Diego, La Jolla, CA 92093, USA

[f] J. Mike Walker '66 Department of Mechanical Engineering, Texas A&M University, College Station, TX 77843, USA

[g] DEVCOM Army Research Laboratory South at Texas A&M University, 575 Ross St (3003 TAMU), College Station, TX, 77843-3003, USA

Lead contact: Dr. Raymundo Arróyave

**Abstract:** Algorithmic materials discovery is a multi-disciplinary domain that integrates insights from specialists in alloy design, synthesis, characterization, experimental methodologies, computational modeling, and optimization. Central to this effort is a robust data management system paired with an interactive work platform. This platform should empower users to not only access others' data but also integrate their analyses, paving the way for sophisticated data pipelines. To realize this vision, there is a need for an integrative collaboration platform, streamlined data sharing and analysis tools, and efficient communication channels. Such a collaborative mechanism should transcend geographical barriers, facilitating remote interaction and fostering a challenge-response dynamic. To further enhance precision and interoperability in this multifaceted research landscape, we must explore innovative ways to refine these processes and improve the integration of expertise and data across diverse domains. In this paper, we present our ongoing efforts in addressing the critical challenges related to an accelerated Materials Discovery Framework as a part of the High-Throughput Materials Discovery for Extreme Conditions (HTMDEC) Initiative. Our *BIRDSHOT* (Batchwise Improvement in Reduced Materials Design Space using a Holistic Optimization Technique) Center has successfully harnessed various tools and strategies, including the utilization of cloud-based storage, a standardized sample naming convention, a structured file system, the implementation of sample travelers, a robust sample tracking method, and the incorporation of knowledge graphs for efficient data management. Additionally, we present the development of a data collection platform, reinforcing seamless collaboration among our team members. In summary, this paper provides an illustration and insight into the various elements of an efficient and effective workflow within an accelerated materials discovery framework while highlighting the dynamic and adaptable nature of the data management tools and sharing platforms.

**Keywords:** Data Management, Materials Discovery, Workflow, Collaboration, High-Throughput Materials Discovery for Extreme Conditions (HTMDEC) Initiative

*Corresponding Author
 Email address: mrinalini.mulukutla@tamu.edu

# I. INTRODUCTION

**Background and Significance of Materials Discovery**

Conventional material discovery processes largely involve trial and error methods along with systematic approaches to identify, develop, and characterize new materials with specific properties. Many transformative material discoveries resulted from the combination of systematic development and happenstance, driven by prior knowledge in related areas. However, the complete materials discovery cycle often requires a considerable amount of time, typically spanning decades, encompassing trial and error to learning and, finally, research and development. These long development cycles and associated high costs pose significant challenges to most technology development efforts. Alberi et al [1], summarized a list of few materials discovery projects to give a comprehensive overview of the current state of the field.

The rise of automation in the past few decades has become a powerful tool in accelerating materials discovery cycle. Early efforts along these lines were concentrated on automating synthesis, processing, and characterization techniques [2,3]. These efforts have not only offered cost-efficiency and eliminated opportunities for human error but also opened new avenues for efficient exploration of the vast and complex materials design space by leveraging emergent toolsets of artificial intelligence/machine learning (AI/ML). The overarching goal of accelerating the pace of materials discovery is possible by integrating automation with artificial intelligence (AI) and potentially enabling autonomous discovery platforms [4,5]. There have been several recent efforts to develop materials discovery workflows in the areas of chemistry and energy storage [6-8]. However, the development of structural materials requires a number of special considerations in developing a workflow that can take advantage of a broad array of experimental and computational tools to effectively navigate an alloy space for a specific set of material objectives. As the scale and complexity of these missions increase, so does the volume of data generated, presenting a vital challenge: how to manage, store, analyze, and make sense of the vast amount of information at our disposal. This fundamental shift towards data-driven materials discovery is revolutionizing how we explore and understand materials and underlying mechanisms. However, the success of these methods hinges on our ability to effectively manage the enormous data sets produced during the discovery cycle [9,10].

In the current work, we present the various elements of an iterative materials-by-design workflow to build the next-generation autonomous materials discovery loops. In addition, this effort describes the data challenges, and solutions adopted to navigate the massive volume of data generated in a large-scale iterative materials discovery mission. By systematically mapping the discovery workflow, identifying challenges associated with each task, levels of automation required, and expanding autonomous loops, we worked toward a more effective, data-driven, and automated approach to transform materials discovery. From data acquisition to storage, analysis, and knowledge extraction, this article offers a comprehensive look at the crucial role data management plays in shaping the future of scientific materials discovery.

**Problem Definition**

The development of structural materials that can operate in extreme environments has a strong potential for improving the reliability and versatility of engineering systems in nuclear, aerospace, and defense industries. One such initiative is "High-Throughput Materials Discovery for Extreme Conditions (HTMDEC), which seeks to accelerate materials discovery by coupling automation and machine learning (ML) techniques to material synthesis and characterization. This mission is focused on advancing materials science through data-driven approaches for developing methodologies, models, algorithms, synthesis and processing techniques, characterization, testing methods, and tools to accelerate the discovery of novel materials. The main goal is the integration of Data-driven Material Design, High-Throughput Synthesis & Processing, High-Throughput Characterization, and ML-augmented Physics-Based Models, along with efficient data handling and management into a cohesive framework to represent a holistic approach to materials discovery.

In this current work, the "Batch-wise Improvement in Reduced Materials Design Space using a Holistic Optimization Technique" (BIRDSHOT) framework was proposed as a highly integrated, interdisciplinary research center focused on accelerating the discovery of advanced structural and multifunctional materials operating in extreme environments. BIRDSHOT brought together a coalition of experts of diverse backgrounds, from geographically dispersed locations spanning across four different time zones, to implement an advanced Bayesian framework. This framework aims to facilitate efficient, optimal, and parallel exploration of intricate materials spaces, and contribute to disruptive foundational research.

Harnessing the power of materials data through an effective data management strategy is a crucial requirement for the success of materials development initiatives such as HTMDEC [12,13]. Data organization across a multi-institutional collaboration (such as BIRDSHOT) must be intentional. Even with the advantage of established relationships between participant organizations and their shared purpose, extensive efforts are needed to create an appropriate schema that would be capable of capturing the relationships between data collection modalities. In this paper, we present several elements that meet the demands of a highly integrated and continuously evolving collaborative research being undertaken in BIRDSHOT.

Collaborative work platforms should enable the capturing and management of comprehensive metadata, annotations, and comments, fostering a culture of interdisciplinary collaboration [14]. Data integration pipelines, supported by machine learning and AI, can fuse diverse data sources for a more holistic materials knowledge base. Adopting interoperability standards, open application programming interfaces (APIs), and version control systems facilitate seamless data exchange and reproducibility [15]. Training and education initiatives empower researchers with data management skills and promote a culture of data sharing, removing geographical constraints and enabling remote engagement and dynamic collaboration in high-throughput materials research. Considering the goals of the BIRDSHOT program led to the following requirements for a data schema.

- Traceability: Measurements must be traceable back through the data management system to the principal control variable, composition, and subsequent process history. Each iteration should be uniquely and intuitively identifiable through a prescribed designation. Once samples of targeted nominal compositions are synthesized, all subsequent measurements, material transfer, or sample sectioning should be captured through a sequence of source-destination relationships that accurately trace the complete chronology and metadata associated with all characterization protocols employed on the samples. Ideally, this will produce a well-defined hierarchical structure.
- Extensibility: A fundamental requirement of a successful data management system is that it is extensible - able to expand over any number of iterations of the control variable or any permutation of the initial synthesis and characterization trajectory.
- Flexibility: The research program objectives were contained within well-defined guide rails that establish a complete materials research trajectory from raw material synthesis (homogenized ingot) to unfinished product (cast or powder production) to finished product (test coupons), wherein each step follows a planned sequence of analysis and characterization steps. However, changes to the control variables and research protocols are inevitable; the data management system should be sufficiently flexible to accommodate such changes.
- Consistency: There is a critical need to maintain consistency across the different operators and equipment used in collaborative research.

## II. MEASURES AND METHODS

The challenges identified so far call for the establishment of standardized data formats, metadata, and measurement techniques, ensuring consistency and clarity. We have concentrated on the following challenge areas over the past year to facilitate the end goal of the project.

- ➔ Focus on designing the iterative and modular workflow: The dynamic and evolving demands of the materials discovery cycle necessitate a systematic and adaptable approach that accommodates the complexities of various tasks involved in development. We have prioritized designing an iterative workflow (i.e., multiple cycles of selecting, synthesizing, characterizing, and assessing properties based on initial design objectives) from the beginning of project activities. By meticulously structuring this framework in a modular fashion, we ensured that team members could efficiently navigate the intricate process. We defined tasks, requirements, and goals within the scope of the project allowing for continuous improvement and incorporation of insights and feedback at critical junctures.
- ➔ Focus on data management and file structure development: We also initiated the definition of data management requirements and objectives right from the onset of the project activities. A data management team, involving members from different domains in the workflow, was identified to enhance collaboration, transparency contributing to the overall success of the project's endeavors. By emphasizing these aspects early on, we laid the foundation for effective data handling and fostered a culture of responsible data stewardship among team members. This data management strategy focused on establishing protocols for data entry, conducting regular data management training sessions for all members of the team, and ensuring openness and accessibility to data throughout the entire materials discovery process.
- ➔ Customized file schema and data collection platform designed to accommodate each step in the workflow: It is critical to align the data infrastructure precisely with the unique requirements of the workflow, ensuring that data can be collected, organized, and analyzed efficiently at each stage. We created a unique file structure and data gathering platform to strategically streamline the data administration and analysis. We organized data in a way that makes sense for each synthesis loop (Vacuum Arc Melting, VAM or Directed Energy Deposition, DED) by modifying the file schema, increasing data accessibility, lowering the likelihood of mistakes or inconsistencies, and facilitating data-centric project execution.
- ➔ Focus on automation of each task in the workflow: We recognize the role played by automation in streamlining processes, reducing manual effort, enhancing efficiency, and reducing human error within the workflow. We have gradually implemented automation for numerous tasks starting with the generation of sample travelers, extending to data collection and preprocessing, to advanced analytics and decision-making to ensure that routine, time-consuming tasks are executed with precision and speed. We intend to adapt to dynamic environments, optimize resource utilization, increase productivity, and stay competitive in an ever-evolving landscape by embracing a more agile and future-ready way of working, where technology works in tandem with human expertise to drive progress and innovation.

In the following sections, we will describe our efforts to create the data infrastructure components that incorporate many functionalities outlined so far. We will examine the complexities of our work to create a strong basis for data management and exploitation, demonstrating the adaptability of our strategy.

**Workflow**

The materials science community has made significant progress in leveraging high-throughput methods and computational techniques to accelerate materials discovery in recent years [16,19,20]. However, the full extent of automation in the closed-loop discovery cycle has not yet been achieved [17,18]. Integrated Computational Materials Engineering (ICME) methods build and exploit process-structure-property-performance (PSPP) relationships, however, they do not readily incorporate extensive data sets from experiments within their framework. Traditional high-throughput (HTP) combinatorial computational and experimental approaches are incapable of dealing with high dimensional and complex materials spaces, and they are 'one-shot' or 'open loop' schemes without a built-in iterative framework to guide future actions, given the current knowledge. Moreover, traditional ICME and HTP approaches do not effectively use (human/time/financial) resources.

Our effort seeks to establish a closed-loop materials discovery and optimization framework in which we integrate all the required elements for accelerated materials discovery combining domain experts from (i) Data-driven materials design; (ii) High Throughput (HTP) Synthesis and processing; (iii) HTP Characterization of static and dynamic mechanical response; and (iv) ML-enhanced physics models of high strain rate deformation behavior, over a targeted materials design space. Information from these four major thrusts is incorporated into the Multi-Information Source Batch Bayesian Optimization (MISBBO) discovery engine [21,22] to provide intelligent suggestions on the source of information to be used at each stage of the process (Fig 1).

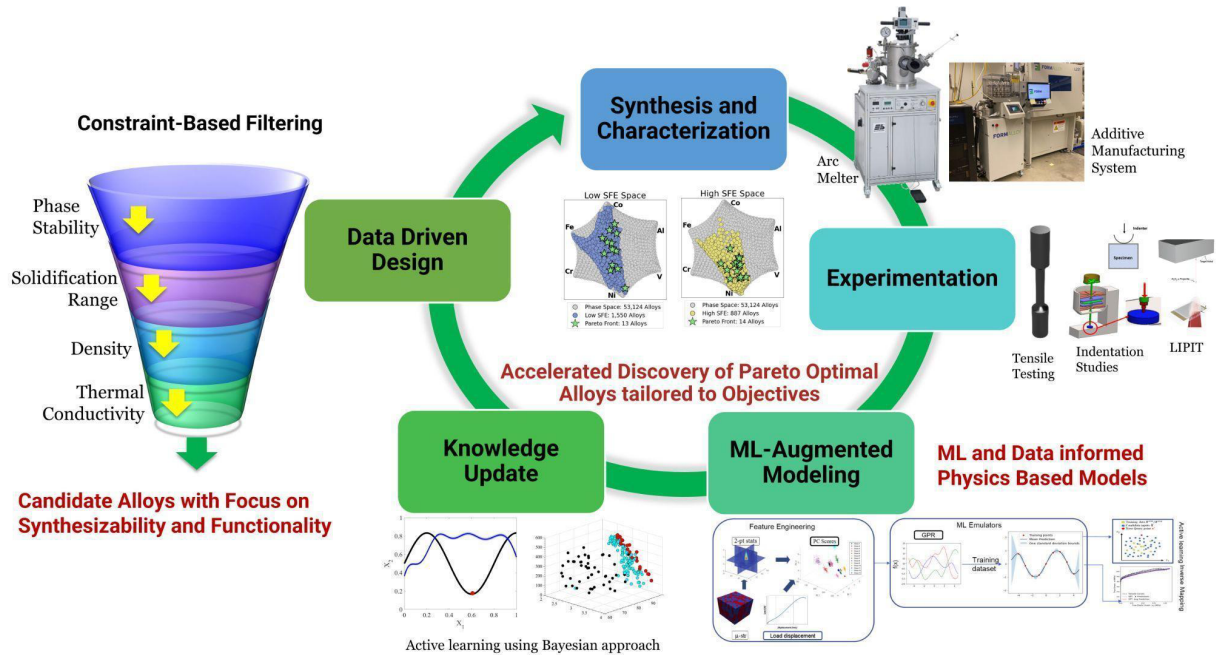

Fig 1. Closed-loop materials discovery and optimization framework designed for BIRDSHOT showing the integration of main thrusts of interest.

In the first year, BIRDSHOT executed each of the thrusts of interest of HTMDEC in an integrated manner. To focus our efforts, we investigated the compositionally complex alloy system, Fe-Ni-Co-Cr-V-Al, to identify the Pareto front corresponding to the maximum strength, hardness, and hardenability to mitigate failure driven by plastic strain localization. The selected scientific problem focused on optimally discovering the multi-objective Pareto set over a vast six-dimensional compositional space. Robust strategies and protocols are established to identify regions in the chosen chemical space to yield synthesizable materials meeting at least some of the minimum performance requirements of interest.

The objective is first to filter the alloy space based on constraints that are relevant to the targeted objectives. This information was, in turn, passed to two parallel synthesis loops, one focused on HTP DED-based sample fabrication and another using HTP VAM. By creating alloy twins under different processing conditions, we aimed to learn the differences and similarities in synthesis routes toward efficient alloy discovery. Samples synthesized in these synthesis loops were analyzed for chemical composition, microstructure, phase analysis and microhardness (MH) testing to determine the properties of candidate alloys. Bulk mechanical properties for samples synthesized by VAM were analyzed by tensile testing. In the synthesis loop by DED, bulk mechanical properties were analyzed by a small punch test (SPT) to accommodate for the sample size limitations of DED samples. Two parallel High Strain Rate (HSR) characterization techniques were used to query the dynamic response of the synthesized alloys. While HSR-Nanoindentation (NI) would test samples at strain rates of up to $10^4$-$10^5$/s, Laser-Induced Particle Impact Testing (LIPIT) would be able to sample behavior at strain rates of up to $10^7$-$10^8$/s. In parallel physics-based simulations, the mechanical response under these two different testing conditions (HSR-NI and LIPIT) would be combined with physics-informed ML models to develop fast-acting surrogate models that would be used to parameterize phenomenological models, which in turn would serve as low-cost information sources to be used to make optimal experimental design choices through the MISBBO approach. Finally, we utilize Batch Bayesian Optimization (BBO) to make globally optimal iterative decisions on where to explore/exploit the design space to maximize the expected gain in performance based on the objectives of maximizing ultimate tensile strength to yield strength ratio (tensile test/SPT), hardness at a specified strain rate (NI) and strain rate sensitivity (NI). This last task closes the loop, enabling the efficient exploration of the multi-objective performance space.

In summary, our effort combines multiple domain experts from computation, synthesis, characterization, and modeling who are located at geographically distributed organizations, collaborating synchronously to perform tasks in a closed loop with the goal of solving a scientific problem. The first successful step implemented in the project was the identification of tasks in the closed-loop workflow, where we clearly defined the roles and responsibilities of the teams involved (Fig 2).

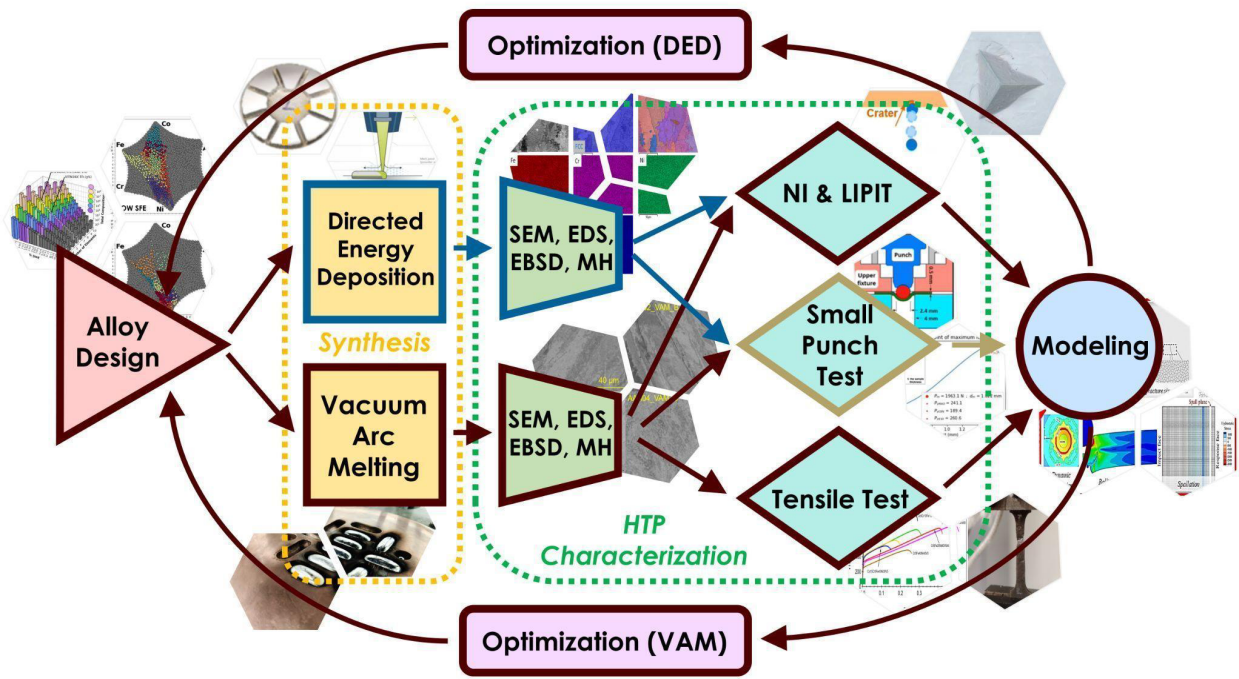

Fig 2 Illustration of a highly integrated and collaborative workflow involved in the BIRDSHOT framework for accelerated materials discovery.

### Challenges and Optimization

BIRDSHOT navigated a complex materials discovery workflow, and various challenges arose throughout the process. To maximize the productivity within the program's timeframe of one year, the initial plan aimed for a 2-month cycle per iteration allowing for five iterations within a year to meet overall project deliverables. This leaves 2 months of buffer for initial planning and reporting, ensuring timely completion. However, encountering bottlenecks threatened the timeline and required strategic adjustments. Some of the challenges encountered and strategies adopted to tackle them are discussed below.

The first month and half of the project was dedicated to establishing comprehensive systems for managing samples, collecting data, and sharing information (naming, folder structure, metadata forms) laying a foundation for efficient data collection and knowledge sharing throughout the project. However, this initial investment of 1.5 months impacted the planned timeline, requiring adjustments to ensure project completion within the defined timeframe.

Each synthesis loop is designed to analyze for composition, microstructure, phase analysis and mechanical properties to evaluate the characteristics of candidate alloys. VAM samples are large enough to allow for sectioning and individual distribution for various planned tests. However, machining and polishing presented a hurdle, slowing down data acquisition, particularly for characterization techniques that require extensive preparation, including LIPIT, Nanoindentation, and SEM/EBSD. To improve the throughput, the team proposed a single-sample strategy for these bottlenecked techniques. This leverages the large sample size by utilizing one sample for multiple analyses, maximizing data output while minimizing preparation time. This solution aimed to improve throughput and address the workflow's complexity.

Another significant bottleneck was posed during nanoindentation testing of candidate alloys due to differing sample mount sizes resulting from the two synthesis paths for VAM and DED samples. Custom sample holders were designed to accommodate the distinct sample sizes, overcoming compatibility issues with standard nanoindentation holders. NI testing to gather strain rate sensitivity (SRS) data initially involved multiple tests at varying strain rates. However, it was observed that this NI testing method produced scattered and unreliable results, including negative SRS values. To address this challenge, a relatively new method called strain rate jump test was employed to achieve more accurate measurement of small SRS values without unrealistic outcomes. Testing metallic alloys also introduced a pile-up phenomenon during indentation, necessitating time-consuming manual area measurements for each of the 144 indents that were planned per iteration. To streamline this process, an automated imaging method was developed. Automation for area measurements is still in development and ongoing efforts aim to address this aspect.

The project's success hinges on the team's ability to adapt and innovate – exemplified by identification and resolution of bottlenecks. This agility is crucial in complex material discovery workflows, where achieving goals within a defined timeframe often requires overcoming unforeseen challenges.

**Weekly Meetings**

In the ever-evolving landscape of materials science, the need for seamless collaboration among researchers, scientists, and engineers cannot be overstated. An integral part of managing the workflow, as well as organizing and coordinating efforts, is regular and open communication between members of the project. Members of BIRDSHOT had a virtual meeting every week, to discuss and plan the activity progress of the entire workflow. These gatherings served as a cornerstone for knowledge exchange, innovation, and coordination among multidisciplinary teams. The team implemented several basic best practices to streamline workflow and enhance collaboration. An agenda is provided before each meeting to help set clear objectives and expectations from each team. Another component was the commitment to starting and finishing meetings on time, or nearly always doing so. Each group reported on their progress during these meetings in accordance with the agenda, proactively shared data, communicated their efforts, and made group decisions on future work, coordinating, and continuously improving the workflow. Meeting minutes are promptly distributed after each session to document discussions and decisions taken providing a reference for future work. Weekly cadence of meetings ensured that groups were focused on the main goal of the project while also investing in finding new ways to perform their tasks more efficiently, as well as continuously looking beyond the main goal to discover new and interesting results. Regular, predictable meetings can enhance communication and maintain momentum in complex workflows. The incorporation of these best practices aligns with the principles outlined in the article. By prioritizing clear communication, punctuality, and regular involvement, the team fostered a conducive environment for successful collaboration and the implementation of the integrated materials discovery framework. While the frequency of meetings was effective for this effort, there is no indication that the frequency, duration, or number of attendees was ideal. It has been shown that regular communication is essential for clarifying ideas, coordinating efforts, and for managing diverse, interdependent research activities [23,24]; however, there are many considerations that should be employed to ensure that meetings are effective in achieving these goals.

**Information & Sample Management**

In the dynamic landscape of scientific materials discovery, where experimentation, analysis, and validation are recurrent processes, the ability to adapt, evolve, and fine-tune our methods is paramount. The iterative nature of BIRDSHOT has been a powerful catalyst in shaping the design of the data and information workflow. As the pursuit of novel materials often demands multiple iterations of data collection and refinement, the data and information workflow must be agile, responsive, and capable of accommodating continuous cycles of discovery [16,25]. Harnessing the full potential of this iterative methodology is only possible with a data management strategy that seamlessly supports capture, organization, and analysis of data at each iteration, ensuring that valuable insights are not lost in the process. In the following sections, we describe our strategies and technologies employed for data management, collection, and extraction in detail.

A. **File System: Google Drive**

With data management being the central part of the research workflow, it has become increasingly important to adopt strategies for consistent and accessible data management practices to meet the goal of promoting open science. Research data management, especially in case of collaborative projects, can be complicated and influenced by a large set of factors making its efficiency crucial for continued research. One of the key challenges of collaborative interdisciplinary projects was the variation in data types and factors involved in data collection at each step. It is imperative for all data and metadata collected to be accessible on a common platform in order to perform holistic analysis and allow for data integration, metadata standardization, accessibility, security, compliance, versioning, and tracking. A number of databases and object-based data storage solutions are available with known advantages in performance, scalability, and search capabilities [14,26]. However, their use was constrained when flexible storage options were needed for data from an ever-evolving research environment with requirements unknown in advance but expected to multiply over time [27-29]. Alternatively, a large number of platforms and solutions that were not databases or object-based data storage solutions have also been developed over the last few years. These solutions can vary significantly in terms of some key aspects including their technological architecture, programming interface, user interface, metadata, coverage, scope, and pricing.

Since the initial project proposal efforts, the BIRDSHOT group had concentrated on establishing data management solutions and strategies. Through Google Apps, we had access to a number of web-based, integrated communication and collaboration applications. The data management portal for BIRDSHOT was planned to be an accessible portal that facilitates long-term Research Data Archive administered by the university. We chose a versatile and adaptable cloud-based file system, Google Drive. This choice was driven by the need to handle datasets originating from various instruments at multiple levels of analysis. Given the diverse sources and continuously evolving data formats, it is essential to adopt a solution that can accommodate both planned and unforeseen changes in data structure seamlessly. Utilization of the Google Drive system played an important role in data collection and management for all these reasons. Its flexibility in accommodating changes in acquisition, analysis and data utilization made it a dependable tool for meeting dynamic requirements of our project. The key advantages of Google Drive include its familiar and integrated interface for data navigation. While other platforms like Dropbox, Box, and OneDrive offer similar interfaces, Google Drive stands out by providing file system integration, individual and service-based API, and automated version control. Google Drive allows version control for file traceability and ensures that erroneously overwritten data can be restored. Access to Google Drive was restricted to the students, postdocs, faculty, and collaborators involved in the project team. Established rules regarding the availability and use of data collected as part of the campaign ensured responsible, secure handling that supports the project's objectives while protecting the interests of all stakeholders. All data and metadata were directly accessible from the file structure standardized for BIRDSHOT. Furthermore, integration of Google Desktop on any machine was a key component of our data management strategy. This feature enabled the online repository to be viewed as a virtual drive from any machine, facilitating automatic data uploads from devices involved in data collection or processing. This capability significantly contributed to the seamless flow of data without creating bottlenecks in the data management process.

Implementing a Google Drive-based file system for BIRDSHOT workflow offered a scalable solution and was helpful in streamlining the research process. By leveraging Google Drive's robust cloud storage and collaboration features, research teams have been able to seamlessly share and access their work from anywhere. This system promoted real-time collaboration, allowing members of BIRDSHOT to work simultaneously, share insights, and provide feedback, thus accelerating the discovery process. While the Google Drive system is not unique in these offerings, it is recommended to adopt a system that provides these features to cater to the diverse demands of complex multidisciplinary projects. After this materials discovery campaign was completed, the Google Drive-based file system contained more than 18,000 files in 3,300 folders, comprising 59 gigabytes of data. This file system was wholly suited to this specific BIRDSHOT workflow, but other considerations should be made for extremely large data repositories that collect many terabytes of data on an annual basis.

B. **Sample naming convention**

The high throughput method chosen for the BIRDSHOT group involves testing a large number of samples, which must be kept organized and identifiable between themselves. Additionally, the chosen organizational protocol must be adaptable, traceable, and easily readable by users [28]. Adaptability allows for the protocol to be modified for new samples, processes, measurements, and investigatory pathways. A traceable protocol ensures sample composition and processing conditions (i.e., thermal, chemical, and mechanical history) are codified, thus allowing for a traceable record of those histories. Finally, the protocol distinguishing samples must be easily understandable by those interacting with it. A 32-digit universally unique identifier, for example, would be the perfect way to separate and provide information about samples. However, these are not conducive to human-powered research settings where sample labels must be intuitive and descriptive, or there is a risk of miscommunication and misnaming.

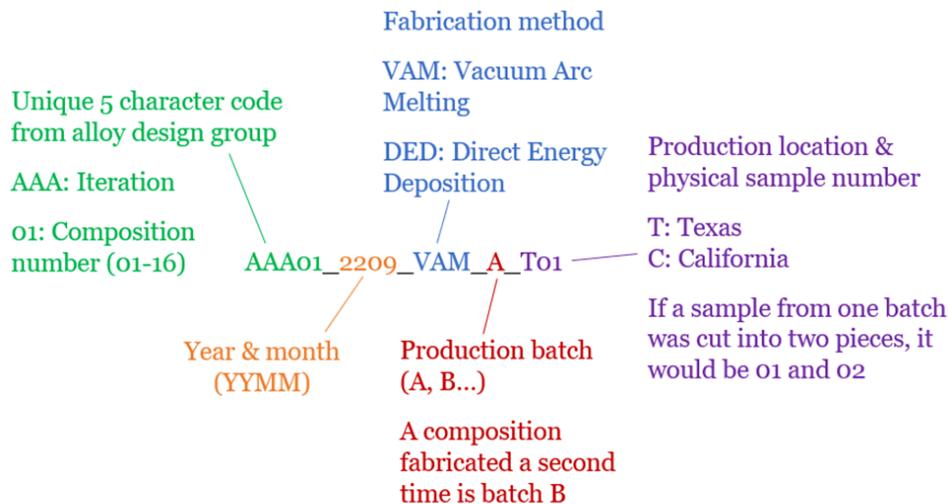

Fig 3. Sample naming convention tailored to BIRDSHOT workflow.

The primary role of the sample name is to provide the reader with a broad context. For one familiar with the project, an effective name will uniquely guide to the appropriate data and will facilitate conversation by providing a proxy for the results and observations associated with that sample. Therefore, the sample name should be compact, information dense, and readily understandable. However, the sample name is not effective at capturing exhaustive metadata and should not be used as such. By uniquely pointing a user to a specific location in the data hierarchy, the sample name should instead point the user to more exhaustive metadata.

A sample naming convention was developed to meet the needs of the center. It was designed and tailored for BIRDSHOT to replicate the workflow and provide enough information to make each individual sample distinct from every other sample (Fig 3). The sample name starts at the beginning of the workflow with a unique five-character code from the alloy design group. The first segment of information includes a three-letter iteration code denoting the iteration number (AAA for the first iteration, AAB for the second iteration, etc.) followed by an assigned composition number from 01 to 16. Subsequently, the date (month and year) of the alloy design in the specified iteration is appended. The fabrication method, indicated as VAM (vacuum arc melting) or DED (directed energy deposition), is listed after the date. This allows the processing history to be known from just the sample name, and new abbreviations can be used if other processing methods are added. The naming convention includes a production batch letter to account for any additional samples, in case the exact same alloy has to be remade for any reason. The final segment of the naming convention indicates the location where the synthesis was performed and includes a sample number if one sample is cut into multiple pieces. This last number in the name is used to identify specific segments cut from the original ingot, which may be tested in subsequent characterization steps (Fig 4). A cutting protocol was designed to standardize the procedure and adopt sub-sample naming conventions.

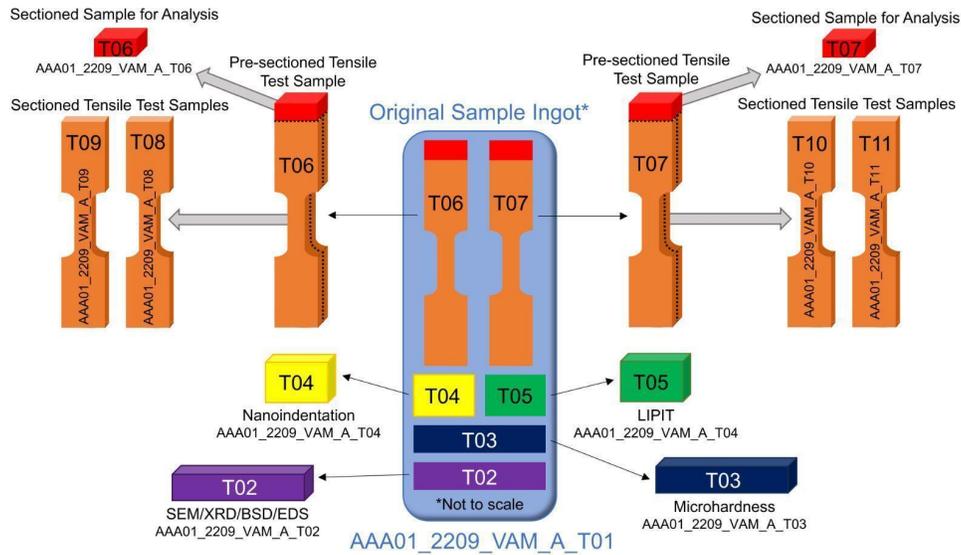

Fig 4. Sectioning protocol for samples synthesized by vacuum arc melting

C. File Structure Standardization

BIRDSHOT requires a large amount of data to be collected, and the utility of data not stored in an intuitive and efficient file structure is severely limited. To be intuitive, the file structure and file names should be self-descriptive and lead the user to the data and corresponding metadata they need. This requires the structure to be workflow-dependent, and, in this case, the structure is dependent on the processing route. The BIRDSHOT team currently employs two different synthesis routes, DED and VAM, with two different yet corresponding sets of characterization steps (e.g. XRD performed at different facilities and using equipment from different equipment manufacturers), which must be reflected in the file structure.

The file structure is designed to address several requirements of an effective data organizational strategy. The structure captures the intentional, probative relationship that exists between specimens as the Original Sample Ingot (Fig 4) is progressively sectioned and subdivided into specimens and test coupons. Sectioning, polishing, and measurements, such as nanoindentation, microhardness, and LIPIT, alter the state or produce a result that is indicative of the original specimen. While many file structures are possible, the structure in Fig. 5 is chosen to ensure properties of specimens further into the file tree are representative of their predecessors up to and including the unique Composition Sample, e.g. AAA01, but excluding the batch, synthesis method, or iteration code. In addition, properties and metadata stored at each node — including steps, specimens, measurements, etc. — are conditions under which the properties of descendent nodes are measured. That is, Tensile measurements from specimens T08 and T09 are conditioned on the properties of Sample T06 which is, itself, predicated on the properties of Sample AAA01, c.f. Fig 4.

This structure makes finding sample data and metadata intuitive if one is familiar with the overall workflow. Similar to the naming convention, the file structure begins with the unique iteration code, which is followed by folders indicating the fabrication method and the batch samples were made in. At this point, the file structure deviates depending on the sample fabrication method. If the sample was fabricated using VAM, the next folders are the individual sample codes (e.g., AAA01) and then the location and sample numbers (e.g., T01) as determined by the sample cutting procedure. For each location and sample number, the sub-folder is the predetermined characterization method that is performed on that sample. If the sample was fabricated with DED onto the eight sample "wagon wheel" holder, then before the individual sample code folder is a sample group folder (e.g., AAA01-AAA08). This particular sample configuration was effective in improving the throughput for polishing, capturing EBSD data, and carrying out other experimental tests on a catalog of 8 samples at a time using methods outlined by Vecchio et al. [30]. The subsequent file structure is essentially the same as VAM but without a cutting procedure that dictates additional location and sample number folders. We considered this wagon wheel schema for all of the samples that were produced by VAM as well. However, the large differences in the mechanical behavior of alloys in a single iteration created difficulties in creating a satisfactory polish for all of the samples in a single run. In some cases, it was possible to apply vibratory polishing for several days to create an EBSD quality surface finish on a large batch of samples, but selective etching of certain grain orientations created a topography that was not suited for nanoindentation.

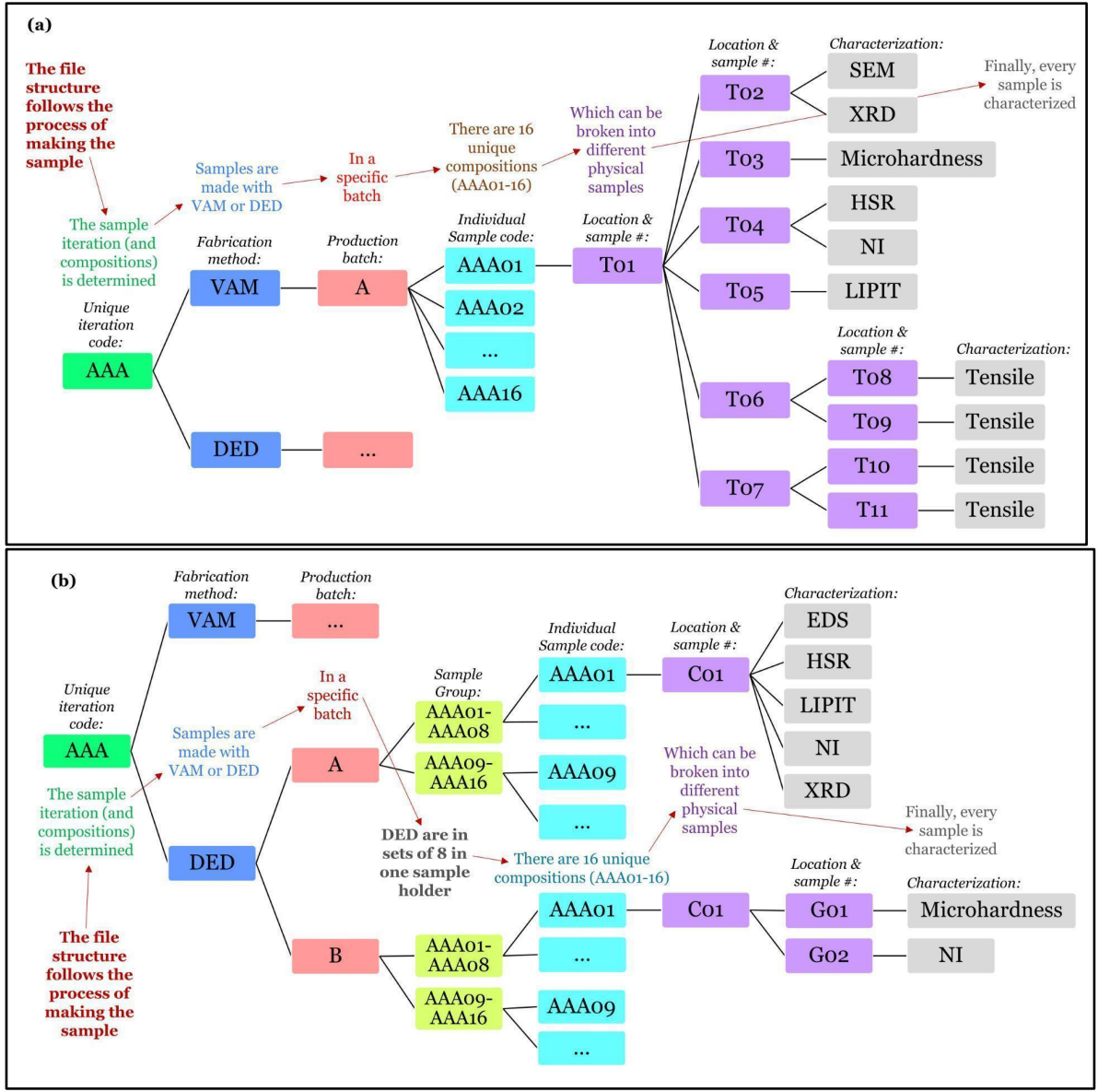

Fig 5. Optimized folder Structure for streamlining materials synthesis workflow via (a) Vacuum Arc Melting (VAM) and (b) Directed Energy Deposition (DED).

This collaboration aims to implement automation solutions that streamline data handling procedures to create user-friendly tools and processes that simplify data input and retrieval, reducing manual effort, enhancing accuracy, and ultimately optimizing our workflow [31,32]. A hierarchical system for metadata creation has been adopted to avoid inconsistencies in data collection. The file structure system allowed for metadata to be stored flexibly and was adaptable to different data types associated with various processes involved in the workflow. The current file structure is reflective of the current workflow (and sample naming convention) with slight differences depending on the synthesis route, thus, anyone can find data and metadata if they are familiar with the workflow and/or the sample name. However, this detailed file structure has been tedious and cumbersome to add and extract data from (an example is available in Appendix A). Data collection in this approach requires going back and forth between multiple folders and files, which can be laborious and prone to mistakes. In addition, the existing file arrangement accommodates all scheduled testing, but lacks flexibility to adapt any supplementary tests that may be conducted on samples in the dynamic materials environment. Recognizing the need for improvement, we are actively developing a more efficient solution.

D. Sample travelers

One of the goals of BIRDSHOT is to accelerate materials discovery by iteratively optimizing every step of the highly collaborative workflow for efficiency, accuracy, and promoting scientific development [33-35]. With intensive collaborative work between multi-disciplinary, cross-functional and geographically dispersed teams, data creation and retrieval play a vital role, and document management becomes a constant uphill battle. The first step in the process of data collection is to address the challenges posed by having a wide spectrum of activities involved in the complex workflow. Specific job traveler templates have been developed for each task of the workflow to capture metadata relevant to the activity and include all necessary data at a convenient location accessible to all stakeholders (an example traveler is included in Appendix B). These travelers are uploaded by individual researchers to the file system within Google Drive when the corresponding task is completed. The metadata captures test conditions, configuration, and process information while the data, both that collected during method execution and from subsequent analyses, are stored separately. Thus, these job-specific travelers separate the storage of data and metadata, increasing the efficiency, flexibility, and ability to compile collected data.

This repetitive process with well-defined and standardized templates is a good opportunity to implement automation to ensure the workflow is as efficient and high-throughput as possible. In an attempt to automate workflow management, we developed a "Traveler Template Generator," which is a job traveler automation tool tailored to the predefined workflow established in BIRDSHOT. The Traveler Template Generator is designed to improve efficiency and reduce errors in BIRDSHOT's data collection and recording processes. Upon suggestion of candidate alloys for an iteration, the tool automatically generates an instance of the folder structure described above and inserts the necessary data collection files, populating relevant fields when available. This simplifies the process of entering and generating files for testing prospective alloys, eliminating the need for additional post-generation organizational tasks. Automating traveler creation reduces the risk of human error and saves time for teams to focus on tasks with higher impact.

The folder structure is described using a JSON-formatted schema document that, combined with a bespoke code for building these hierarchical structures starting at any level, improves traceability and repeatability from iteration-to-iteration. The JSON schema supports single inheritance to allow for fully polymorphic entries. Polymorphism is of particular importance as it allows complex tests and processes to be composed from simpler entries, ensuring consistency and comparability between objects. A simplified example is provided in Table 1. Because the underlying file system is hierarchical, additional objects may be inserted at any location in the tree. This allows single-call construction of the "typical" iteration — that is, all the steps, samples, and tests expected in a single iteration — while still allowing additional measurements to be added as discovery dictates.

Table 1. JSON schema demonstrating single inheritance (base) and polymorphism: TestA extends Test and TestB extends TestA. A single command, build --schema simple.json Root Iter, constructs the folder structure on the right. A second command, build --schema simple.json --where "Iter/Sample

| JSON Schema | Folder Structure |
|---|---|
| ```json
[
  {
    "type": "Root",
    "base": "Folder",
    "contains": {
       "Sample A": "Sample",
       "Sample B": "Sample"
    }
  },
  {
    "type": "Sample",
    "base": "Folder",
    "contains": {
       "alpha": "TestA",
       "beta": "TestB"
    }
  },
  {
    "type": "Test",
    "base": "Folder",
    "contains": {
       "Setup": "Folder"
    }
  },
  {
    "type": "TestA",
    "base": "Test",
    "contains": {
       "Results": "Folder"
    }
  },
  {
    "type": "TestB",
    "base": "TestA",
    "contains": {
       "Alignment": "Folder"
    }
  }
]
``` | ```
Iter
├── Sample A
│   ├── alpha
│   │   ├── Results
│   │   └── Setup
│   └── beta
│       ├── Alignment
│       ├── Results
│       └── Setup
└── Sample B
    ├── alpha
    │   ├── Results
    │   └── Setup
    └── beta
        ├── Alignment
        ├── Results
        └── Setup

17 directories, 0 files
``` |

B" TestB gamma, would construct an additional test of type TestB (not shown).

Such consistency is critical because relationships are implicit in the folder hierarchy. The readability and foundation in a text-based description of the structure, allowing for version control, improves traceability, repeatability, and auditability. JSON is also used as an underlying technology to enable validation and consistency in metadata collection. A more fulsome discussion on this is reserved for Section Collections.

With only minimal training and introduction, the resulting folder structure produces an unambiguous location to persist results in the myriad file formats they adopt: CSV, DAT, Excel, HDF5, sqlite, etc. These files may require specialized or proprietary software to process. Others may be amenable to alternate analyses not part of the original project scope. The choice to persist the data in this structure encourages the development of custom analysis pipelines (workflows) that act as needed, on demand, when data is accessed. These workflows are interrogable, traceable, and version controllable, improving transparency in research. And following the well-established software principle of Separation of Concerns, such workflows more clearly delineate collection, curation, and analysis.[1]

The creation of sample travelers for each task has been instrumental in ensuring the seamless progression of research. Sample travelers serve as dedicated documents or records that accompany samples throughout the workflow and an essential communication tool. They not only aided in maintaining transparency and traceability but also provided a comprehensive record of the research journey, which can be invaluable for quality control, troubleshooting, and data analysis. Travelers facilitated effective cross-team communication and collaboration, thereby optimizing the materials discovery process while upholding rigorous data management standards.

E. Sample tracking

BIRDSHOT involves the movement of samples from one geographically distant group to another, as defined previously, and thus results in a sample management challenge. Effectively managing samples as they move between groups requires a simple way to report any and all sample movement. In the preliminary stages of this project, a designated email account was created to track sample movement. When a sample or group of samples is moved between groups, the sending and receiving groups message this designated email address confirming the sample movement. One student is tasked with monitoring the sample tracking email account to ensure no samples are lost and that all samples are where they need to be.

With a workflow that involves a large number of samples, sample location tracking is imperative, so no samples are lost or misplaced. The current sample tracking method involves groups messaging a designated email address to update the sample location manually when samples are sent out or received. Although this method works, it could be modified to be more HTP. BIRDSHOT is currently working with all groups involved in the project to develop a new HTP and easy-to-use sample tracking method. One potential solution would be to include scannable QR codes on samples that automatically update sample location. Additionally, another prospective solution under consideration is the adoption of a project management platform. Though it has not been implemented to date, it holds promise as a valuable addition to the overall project.

F. Knowledge Graphs

The process of discovering new materials with desired combinations of properties requires a large volume and variety of background knowledge. For example, we may want to know how a process changes the structure of a material, each of the steps in that process, and the accepted descriptions of the structure in relation to other materials literature. Therefore, we are adopting a knowledge graph (KG) framework for representing the knowledge captured by the hierarchical file structure. This framework uses ontologies, a type of vocabulary for representing knowledge in a formal mathematical logic, and the resource description framework (RDF), a declarative programming language for digitally representing the vocabulary. All of the

---

[1] This separation between collection, curation, and analysis does not preclude users from persisting often calculated results but shifts the paradigm from retrieval of static data to one of dynamically calculated data.

keywords in the folder structure have a corresponding ontological representation, and each real folder and file name populate the values.

By maintaining a strict conditional–response hierarchy in the file structure (Fig 5.), one that reflects the relationships intentionally imposed between experimental artifacts, properties of specimens found later in the hierarchy are representative of their predecessors and properties of earlier specimens represent the conditions under which descendent properties are observed. As a result, traversal of the file structure, a deceptively simple tree graph, corresponds to knowledge transfer between artifacts.

Aggregation, collecting information from a node to its predecessor, is valid when the properties of the successor are representative of the predecessor; a condition that was imposed by the file hierarchy. If initiated by the predecessor node — e.g., the active node requests information from its successor— then a preorder traversal of the tree will collect properties from each node into its immediate predecessor, and only its immediate predecessor. In contrast, post order traversal will collect into the root of the traversed subtree the properties of all of its descendent nodes.

This latter approach, aggregation through post order traversal was used to collect the properties of all representative samples into the Composition Sample, e.g. AAA01, in Fig. 5. From there, a central tendency, such as the mean for numerical and mode for categorical data, is used to reduce the collected list of semantically equivalent properties[2] into summary statistics.

Analogously, propagation disseminates information from a node to its immediate successor and, as with aggregation, the outcome during propagation depends on the traversal order. In post order traversal, under the condition that the active node pushes information into its successor, data is propagated only to a node's immediate successor. In contrast, propagation under preorder traversal will scatter the properties of a node to all of its descendants.

Brought together, data in the folder structure is collected (aggregation) to the Composition Sample, reduced through a central tendency filter, and scattered (propagation) from the Iteration, see Fig. 5. The two major differences between the aggregation and propagation steps are that, during aggregation, semantically equivalent properties are collected into a list before reduction. During propagation, semantically equivalent properties are skipped. Ontologically, this is founded in the logic that a property existent to a sample is more closely representative of that sample than that property inherited from its predecessor. Colloquially, the existent property is closer to the source.

The second difference is that aggregation stops at the Composition Samples. This seemingly innocuous trait belies a hidden complexity addressed intentionally and handled uniquely within this data management infrastructure. Both data and the algorithms and data processing pipelines required to process the data are part of the data management system. The aggregation step presupposes knowledge of the relationships captured within the folder structure; a knowledge unlikely found beyond the immediate project participants. Therefore, an analysis pipeline (workflow) is developed that defines this three-step data access protocol.

One might reasonably question why this information is not simply persisted in a data store. The reasons are twofold. First, the data are dynamic with additional iterations, batches, and compositions possible should this research focus be restarted, and these self-same algorithms would be required to composite the data collected

---

[2] Because collection and curation of these data are coordinated, consistent semantics were used throughout. Should this approach be applied to uncoordinated data, such as data composited from literature, an additional effort to define, construct, and impose consistent semantics would be required before aggregation and reduction.

during these subsequent investigations. Second, several assumptions are present in the implicit relationships captured by the file structure presented in Fig. 5.

The data hierarchy assumes and relies on descendent nodes being representative of their successors and on predecessor nodes' properties being representative of the conditions under which the properties of descendent nodes are observed. However, future processing steps — such as annealing, case hardening, etc. — that might be applied to interrogate the response of residual samples would require additional consideration to include and incorporate these prospective measurements into the current corpus. Although not currently planned, a sufficient data management system must be designed to handle these eventualities and others that arise from discovery that occurs during research.

This representation is flat and allows direct access to each of the elements, allowing flexible declaration of new relationships between data elements and other information making it ideal for integrating different data sources [36]. Since the field of materials science is very complex, and the process-structure-property (PSP) space is only partially known, a flexible structure is critical for tasks such as HTMDEC that collect an ever-changing and growing corpus of knowledge. Knowledge may be recorded in research papers, handbooks, other technical reports, online repositories, and some databases. Additionally, this project aims to go beyond capturing only the positive results published in papers and books, and much of the trial-and-error information from notebooks and personal records must also be captured. The problem of having these many sources is known as fragmentation of data, where data is not organized so that related pieces of information are accessible together.

An associated problem is the heterogeneity of the data, the differing structure and type of the data, which arises naturally from the fragmented generation and collection efforts. Therefore, choosing the KG technology for representing the collected data is essential for ensuring that the relationships between the fragmented and heterogeneous data sources in this project can be represented. It also enables the integration of these data sources using established strategies such as ontology-based data integration (OBDI) [36]. KG tools, for querying data and inter-relationships between the data and reasoning about the data to discover new relationships [37,38], provide the insight necessary to collaborate at scale and iterate experimentally Applying KGs to materials science requires the definition of an ontological model that describes the materials domain. The correct ontological model is based on the terms in the structure folders as well as an ever-growing definition of background knowledge from heterogeneous data sources; multiple SEM images, their magnification, the phases present in each image, crystal structure and properties of the phases, the samples' processing history, related finite element simulations, and trained machine learning models are examples of heterogeneous types of data needed to describe a particular material system and PSP system of interest.

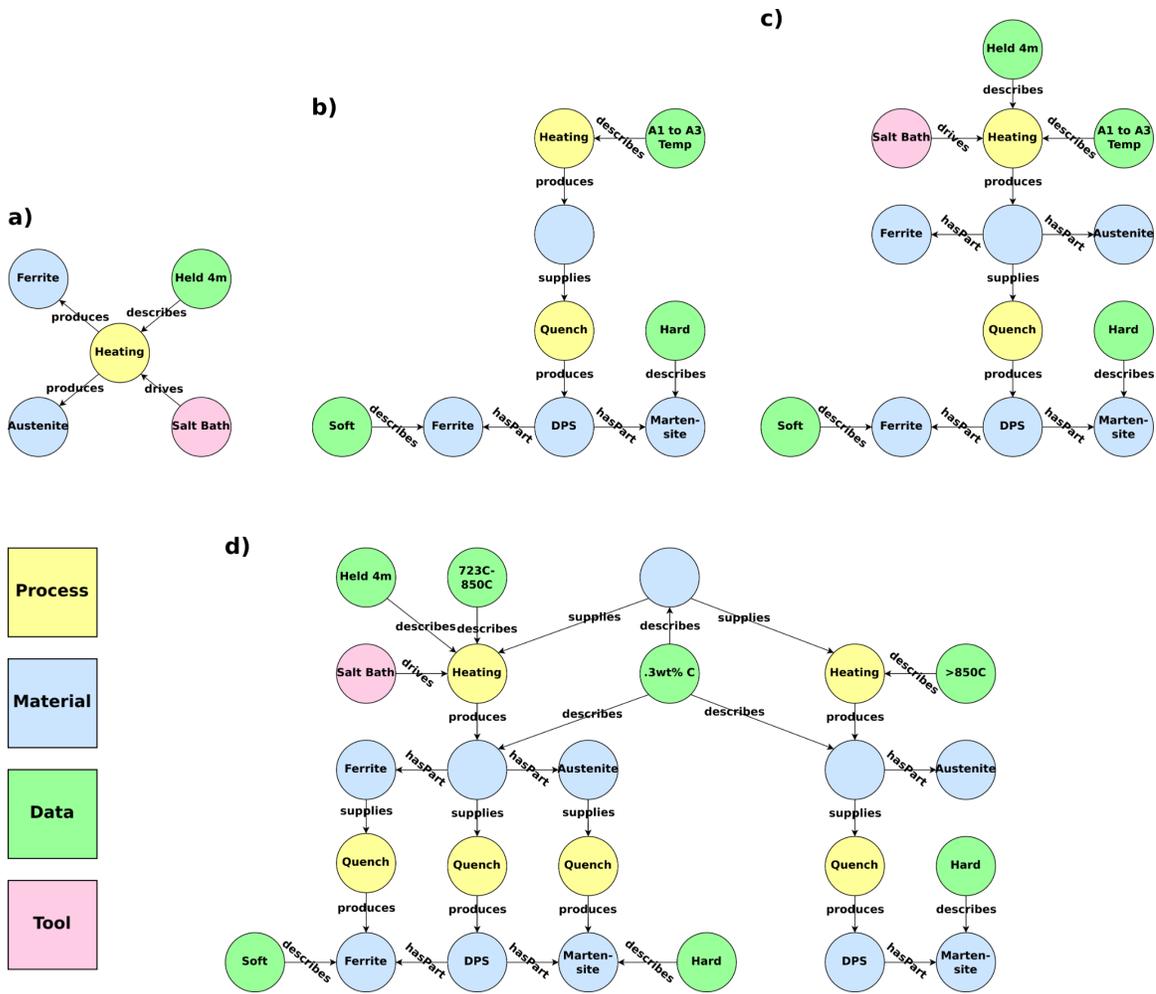

Fig 6. An example KG for dual-phase steel (DPS) that serves as a common structure for aggregating and integrating collected knowledge. In this example, the KGs in a and b are disparate sources of knowledge that follow that same epistemological structure, allowing them to be integrated into the KG in c. As more data is collected and added to c, the KG in d can be generated, which represents additional microstructural information and multiple processing paths for different materials that can be generated from the same initial material. Through the continuous process of data collection, the KG grows and serves as a single, accessible source for users.

Integrating a knowledge graph into the BIRDSHOT workflow is an ongoing effort to incorporate KG insights for unlocking the full potential of interdisciplinary research. The existing effort has defined the standardized folder structure keywords in the RDF format, and follow-up efforts will directly map folder and file names to the KG structure. This would serve as an intelligent framework that captures and connects complex relationships among materials, properties, and processes, transcending traditional data silos. It would help members of the team to make data-driven decisions, discover hidden patterns, and accelerate the identification of novel materials. Efforts are underway to create a specialized knowledge graph tailored to the intricate BIRDSHOT workflow. Given the complex nature of this workflow and the tremendous wealth of background knowledge, the development process is ongoing. This ongoing development signifies a commitment to advancing research in the BIRDSHOT approach, ultimately paving the way for groundbreaking discoveries and innovations in this complex domain.

# Collections

**A. File Schema & Data Collection customized based on process/characterization involved**

A number of data schemas have been developed for materials data, schemas that are intricately connected to the database in which the data are contained. Following a prescribed schema improves data findability, accessibility, interoperability, and reusability (FAIR). Both public and private research has focused on the creation of a universal data standard [39]. Despite these efforts, many factors, both technical and practical, have and continue to inhibit the development of a universal materials schema. Permissive data schema places minimal restrictions on the content, allowing each database implementation to categorize specific material properties into discrete bins [40,41]. Though this maintains internal consistency, it results in inconsistencies and incompatibility across different data resources. Conversely, prescriptive data schemas stringently delineate structures and vocabulary to authenticate data. However, this comes at the expense of narrowing the range of materials subdomains the schema can accurately represent.

Efforts to harness the FAIR attributes of certain schemas have spurred initiatives in materials vocabularies, semantics, and ontologies. Drawing inspiration from the semantic web, which aims to detail the internet's content with clear definitions and established relationships, these ventures seek to precisely define materials data [42-45]. The effectiveness of this semantic approach remains uncertain due to the vastness of both the global web and the materials-specific domain. Yet, whether it's the permissive data schema, the prescriptive one, or ontologies, all these approaches bring structure to materials data. With structure comes the potential to interlink and compare related data.

In their materials discovery and optimization effort, the BIRDSHOT center represents a microcosm of this global challenge. Each group within the organization represents a unique subdomain and produces data that is supplementary and complementary to the data produced by the other groups. To facilitate data interoperability, each group establishes a schema specific to their data streams. A shortened example of a schema is shown in Table 2. The complete schema is available in Appendix C.

While the concept is, itself, technology agnostic, JSON Schema was selected to describe the structure of the data for its ubiquity, its relative simplicity, its flexibility, and because it is a text-based format, the ease with which these schemas can be version controlled and traceable, ensuring that changes are documented and tracked. The elegance, completeness, or capabilities of a schema are meritless without compliance. Therefore, the first and most important step beyond the creation of the schema is the creation of a simple user interface that facilitates the collection of compliant data. The ubiquity of JSON Schema simplified the development of a web application, simply named *Collections*, that parses the JSON Schema into interactive forms. These forms ensure compliance by explicitly presenting fields for expected data and validating that data for type, content, and completeness. Such an interface is simple to set up within a protected enclave but to enable engagement across participating organizations, Collections must be hosted as a public application and, therefore, protected by credentialed access.

While beyond the scope of this initial deployment, Collections must also support multi-tenancy, allowing individuals involved in multiple data collection efforts (projects) to easily switch between projects and be presented with project-specific content without exposing projects to which a given user has no access.

| Schema |
|---|
| {<br>    "properties":<br>        "Sample ID":{<br>            "errorMessage":"Please enter a Sample ID such as AAA01_2209_VAM_A_T04.",<br>            "format":"custom",<br>            "pattern":"([A-Z]{3}[0-9]{2})_(22\|23)((0[1-9])\|(1[0-2]))_(VAM\|DED)_([A-Z]+)_(T04 \|C01)",<br>            "Type":"string"<br>        }<br>        "Indentation Date":{<br>            "description":"YYYY-MM-DD",<br>            "errorMessage":"Please enter a date in the YYYY-MM-DD format",<br>            "format":"custom",<br>            "pattern":"^202[23]-((0[1-9])\|(1[0-2]))-((0[1-9])\|([12][0-9])\|(3[0-1]))([ -](([01][0-9])\|(2[0-3]))(:[0-5][0-9]){1,2})?$",<br>            "type":"string"<br>        }<br>    },<br>    "title":"Nanoindentation (SRJT) Details",<br>    "type":"object",<br>    "description":"Form to track the nanoindentation details for the strain rate jump test (SRJT) method."<br>} |

Table 2. Example of shortened schema demonstrating data interoperability.

B. Integration of code for data analysis

In addition to improving compliance, completeness, and data quality, centralization of data collection has the added benefit that arbitrary analysis pipelines may be set up and triggered as data are collected, and forms are submitted.

Several characterization techniques have been employed in the search for alloy compositions with improved mechanical performance, as defined by the initial program objectives. In this effort, the targeted objectives were to maximize: (1) hardness, (2) strain-rate sensitivity, and (3) the ratio of ultimate tensile strength to yield strength. However, the workflow techniques are highly general, and alternate objectives may be used

depending on the specific materials design problem. Once the methods and means of extracting specific data from each technique are well defined, automation is desirable to improve consistency and traceability across measurements and to streamline asynchronous data synthesis pipelines.

The quasi-static testing method is utilized to evaluate the tensile characteristics of candidate alloys. The force–displacement data produced by this technique is transformed to the familiar, qualitative engineering stress–engineering strain curve and, for this specific analysis pipeline, further transformed to true stress–true strain. Motivated by an eventual move to high throughput, semi- to fully-automated testing and to provide testability and traceability of the transformation and interpretation of this data from its raw force–displacement to its final true stress–true strain form, a bespoke analysis code was written to transform and extract scalar descriptors from tensile data. In a similar fashion, data was extracted from nanoindentation strain rate jump tests, performed using the methods of Maier et al. [46]. Automating the execution of this data both ensures that the transformation has occurred but also provides an opportunity to inject metadata that further improves traceability: execution time, transform software version, user information, and any other information deemed valuable.

Automating asynchronous data synthesis pipelines is common in materials research [47,48]. Data collected from two or more characterization modalities may be combined to construct a composite performance metric or as input to inform an iterative research pathway- both of which are used in the current research paradigm. Analysis pipelines are triggered automatically to summarize hardness, strain-rate sensitivity, and tensile measurements. These place the candidate alloy inside the performance evaluation space. Once a batch of alloy candidates is examined, the locations within the performance space of all candidates, including the most recent iteration, are used to establish the Pareto Front. Those compositions sitting near the Front and their performance characteristics are included in a recommendation engine to suggest the next iteration of candidate compositions.

This data collection model not only separates collection from analysis but also from curation. Automated asynchronous data and analysis pipelines enable the construction of composite performance metrics and iterative research pathways. This approach allows for custom app development and domain-specific workflows, including ETL pipelines for seamless data integration. Additionally, containerization using Docker facilitates the deployment and reproducibility of the data integration environment. This streamlined approach to data collection, analysis, and integration, as illustrated in Fig. 7 enables a more efficient and centralized workflow for materials research, promoting compliance, completeness, and data quality. From raw material production and characterization through component level testing of semi- and finished-goods, data collection and usage patterns form a complex dependency network where data generators and data consumers—metaphorically represented as on- and off-ramps, respectively, in Fig 7—contribute to, access, and even create private, shared, and public data resources Holding to the transportation metaphor, these data resources are local roads, state highways, and interstates.

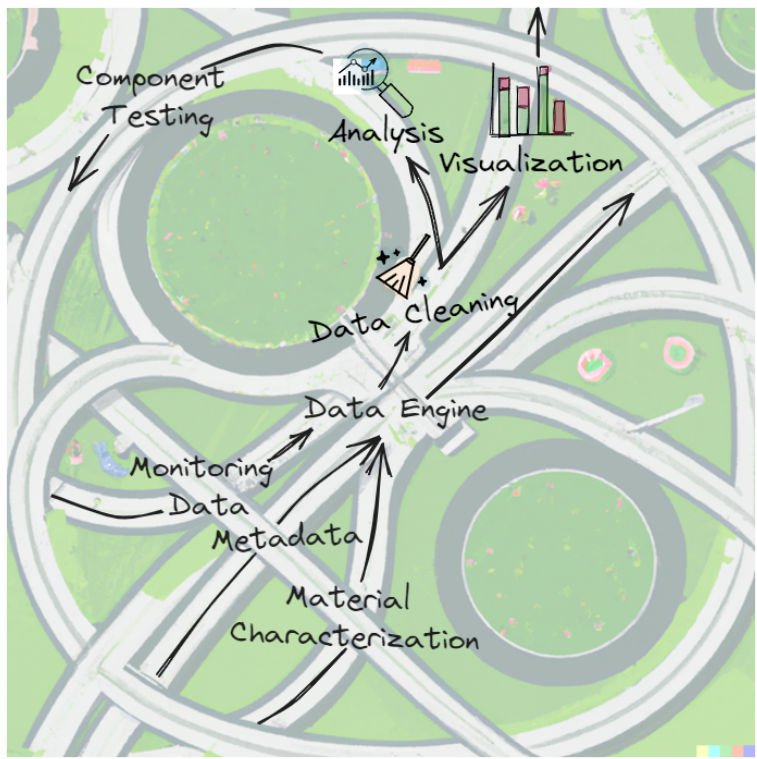

Fig 7. An illustration of a centralized data workflow in materials research, where monitoring data, characterization, and associated metadata channels merge into the main data engine highway with data cleaning, visualization, and analysis off-ramps at varying points in the highway as needed. The analysis off-ramp further branches into component testing, creating a seamless loop for data and metadata processing, iteration, and decision-making points. Background image generated by DALL-E 2.

## III. Integrating Data and Bayesian Insights for Materials Discovery

**Data Management Implementation Timeline**

The progression of Accelerated Discovery through the robust workflow elements in BIRDSHOT has been a notable development, influenced by a consistent focus on scientific rigor. (Fig 8). The project began with the goal of improving research productivity by refining workflow elements, particularly workflow design and optimization. In its initial phases, the emphasis was on creating effective information-sharing and management systems to promote collaboration among researchers, regardless of their location. As the project evolved, it explored data collection and extraction, leading to the development of a modern collection platform. This platform, combined with the simultaneous creation of a detailed knowledge graph, became central to our approach, ensuring data integration and access to key insights. With each milestone, the initiative has changed the research process and set the stage for future advancements in scientific discovery, showcasing the role of technology, data, and teamwork in expanding our understanding.

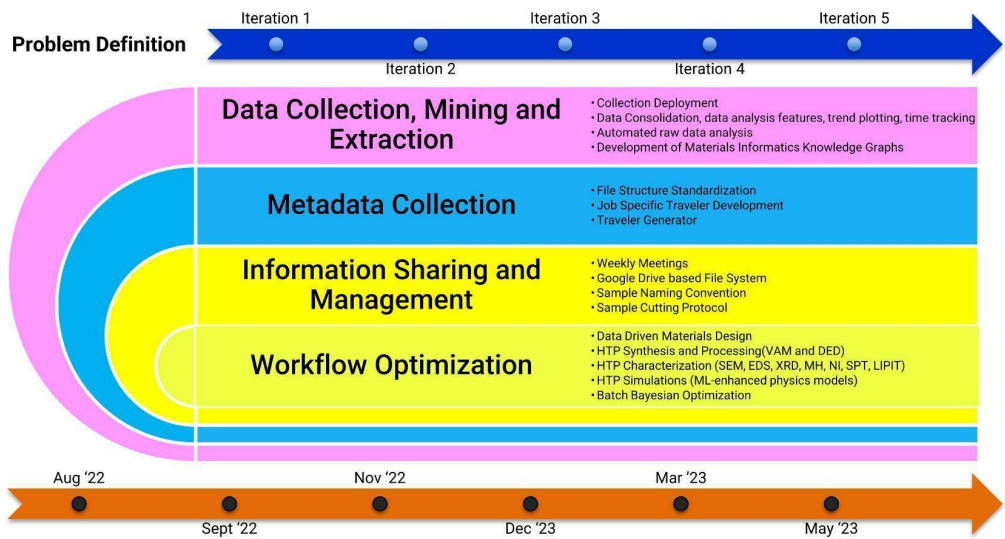

Fig 8. Timeline and Iteration Progress of Accelerated Discovery through Optimization of workflow elements over the past year of project BIRDSHOT, Highlighting Iteration and Month-by-Month Implementation Details

**Leveraging Information Instructure for Bayesian Discovery**

The Accelerated Discovery campaign was guided by an Artificial Intelligence framework based on Bayesian Optimization. A Bayesian discovery campaign allows for updating beliefs in light of acquiring new observations that in the first place are selected strategically to help with identifying high-performance alloy regions [21,22]. In any iteration, the Bayesian design framework, as the decision-making center, searches for experiments with the highest potential to improve the target performance metrics by evaluating a utility function. Such a data-driven materials design approach requires leveraging information to and from different workflow elements to the decision-making center.

In BIRDSHOT, the communication between the decision-maker and experimentalists is established by iteratively transferring proposed experiments to experimentalists and transferring back the experimental data using the developed data collection platform. The Bayesian design framework integrates experimental data as new information to update its state of knowledge and to decide on the selection of the next experiments accordingly. Specifically, the first stage of the project deployed a constraint-aware multi-objective batch Bayesian Optimization (MOBBO) algorithm capable of suggesting multiple parallel queries to the feasible alloy design space. As mentioned above, the objective of this first discovery campaign was to identify a three-objective Pareto set over a multi-dimensional alloy space. The Pareto set consists of solutions with no superiority over each other as improving one objective may deteriorate the performance of other objectives. Fig. 9 presents a schematic of the framework as well as the results of the iterative framework. The Pareto set for this alloy system was discovered in five iterations by querying only 80 alloys out of a design space consisting initially of 50,000 potential candidate alloys. Compared to a traditional Edisonian approach to discovery, this represents a ~100 to 2,000-fold acceleration of the discovery rate—the range varies depending on whether the Edisonian query is carried out sequentially or in a parallel fashion.

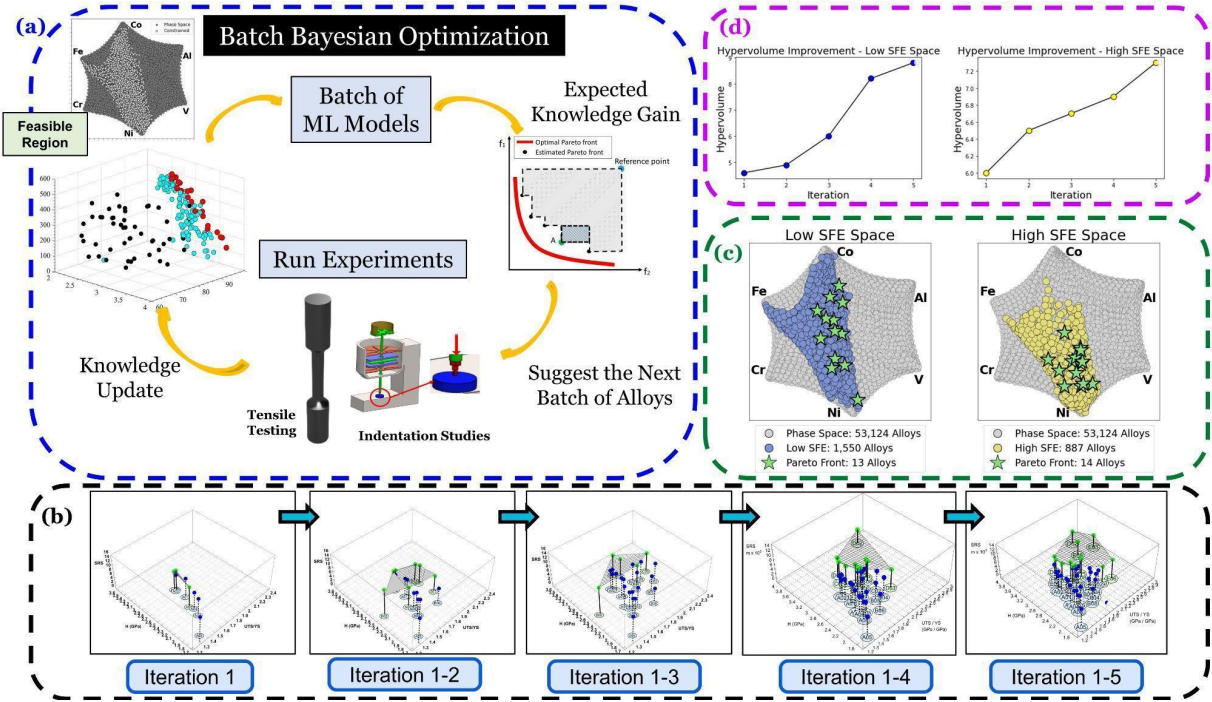

Fig 9. (a) Schematic of Batch Bayesian Optimization framework employed in BIRDSHOT, (b) the evolution of the Pareto front for a multi-objective optimization problem across five iterations, where three objectives are maximized in parallel for a design space consisting of 50,000 candidate alloys. (c) UMAP Representations of Pareto Set derived from the Feasible Space under Intelligent Constraint Filtering and (d) Hypervolume Improvement across five iterations.

## IV. CONCLUSION

A large part of material science revolves around the discovery and implementation of new and interesting materials. This coincides with the goal of the HTMDEC project for rapid alloy discovery in a high throughput manner. Any form of alloy discovery involves multiple fields of expertise, from alloy design and synthesis to modeling and optimization, and the BIRDSHOT center is no exception. To perform such discovery in an efficient and high throughput manner requires an effective data management system and a collaborative work platform that eases the storing and sharing of data. In this work, the BIRDSHOT center has been able to address the most important challenges in Data Management and collection as an integral component for the development of a Materials Discovery Framework, and results are presented. The BIRDSHOT center has been able to integrate the use of Google Drive, a sample naming convention, a file structure standardization, sample travelers, a sample tracking method, and knowledge graphs as data management tools and Carta Collections as a collaborative work platform. Importantly, BIRDSHOT operates with a well-defined workflow and weekly meetings to ensure important information is shared between members. Integration among tasks was driven by a Bayesian iterative approach that has already demonstrated its effectiveness at solving challenging alloy discovery problems. However, the data management tools and data sharing platform used by BIRDSHOT are not stagnant. As members generate and process new data, more specifications and requirements can be realized and subsequently translated into iterative updates to the existing tools and data-sharing platform.

## V. FUTURE OUTLOOK

The BIRDSHOT center has already implemented several elements of collaborative research data management successfully within a short period of one year. The positive experience and outcomes observed upon streamlining tasks as required with researcher involvement emphasize the importance of a well-designed organizational and conceptual approach to research data management. Such approaches are highly recommended to advance data management practices in research projects. This ongoing effort will proactively address the missing parts while refining and improving existing aspects of the overall workflow schema.  Some specific areas of improvement are outlined below:

- Continued streamlining of metadata creation process through accessible electronic notebooks in laboratories to minimize manual entry efforts by researchers.
- Continued efforts in creating tools for automated raw data analysis within the data collection platform.
- Implementation of a meta description framework for automated generation promoting efficiency, consistency, and maintainability. It can be particularly valuable in scenarios where the application's internal model objects frequently change or expand, as it enables rapid adaptation of user interfaces and database structures to reflect these changes.
- Development of data visualization routines tailored to communicate complex information related to targeted objectives.
- Project management software could be used as an improvement to the current sample tracking system that was employed in this study.  This software could also help to improve scheduling of workflow activities, especially in the categories of material synthesis and HTP characterization.
- The automated uploading of raw experimental data to Google Drive was not utilized by all research groups, and efforts will be made to improve this practice for archival purposes.

## Acknowledgments

Research was sponsored by the Army Research Laboratory and was accomplished under Cooperative Agreement Number W911NF-22-2-0106. The views and conclusions contained in this document are those of the authors and should not be interpreted as representing the official policies, either expressed or implied, of the Army Research Laboratory or the U.S. Government. The U.S. Government is authorized to reproduce and distribute reprints for Government Purposes notwithstanding any copyright notation herein.

The authors would also like to acknowledge Texas A&M University's Technology Services for their support in providing Google Apps and services, which have greatly contributed to the success of project activities.

## Compliance with Ethical Standards

**Conflict of Interest:** On behalf of all authors, the corresponding author states that there is no conflict of interest.

## Figure Availability

High-resolution images of figures included in the article are provided as separate files.

## Code Availability

Code for automated job traveler generation, *Traveler Generator*, is available on GitHub at https://github.com/Tripp5118/TravelerGenerator/.

# Appendix-A

Example folder structure for hypothetical samples (a) AAA01_2209_VAM_A_T02 and (b) AAA01_2209_DED_A_C01

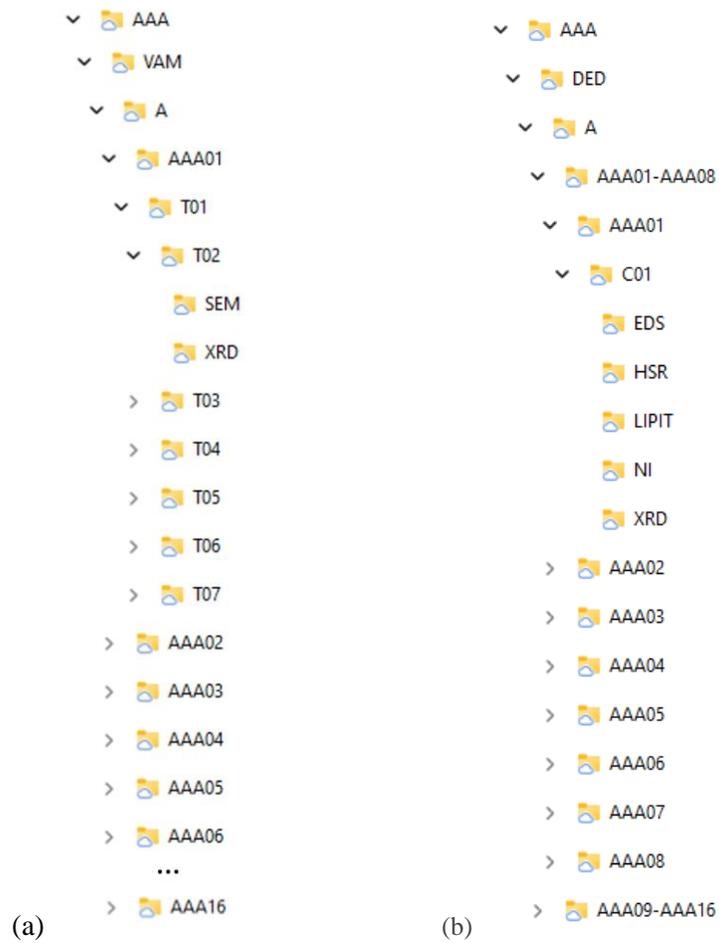

(a)         (b)

# Appendix-B

Example nanoindentation job traveler for performing strain rate jump testing on samples and collecting all relevant data

**HTMDEC Nanoindentation Sample Traveler**    Traveler # _____ of _____

Sample Name: _______________________________________________

Operator: ____________________________   Indentation Date: _______________

Indenter Tip #: _______________________   Tip Cal Date: _________________

# of Indents: ________________________   Device Used: _________________

Time Spent: _________________________   Sample Poisson Ratio: __________

| Machine Settings | | Test Parameters | |
| --- | --- | --- | --- |
| Target Load [mN] | | Initial Depth [nm] | |
| Target Depth [nm] | | Target Depth Jump 1 [nm] | |
| Target CSM Frequency [Hz] | | Target Depth Jump 2 [nm] | |
| Surface Approach Velocity [nm/s] | | Target Depth Jump 3 [nm] | |
| Target Dynamic Displacement [nm] | | Target Depth Jump 4 [nm] | |
| Hold Max Load Time [s] | | Final Depth [nm] | |
| Surface Approach Distance [nm] | | Initial Target Strain Rate ($\dot{P}/P$) [s$^{-1}$] | |
| Data Acquisition Rate [Hz] | | Jump 1 Target Strain Rate ($\dot{P}/P$) [s$^{-1}$] | |
| Depth To Start Averages [nm] | | Jump 2 Target Strain Rate ($\dot{P}/P$) [s$^{-1}$] | |
| Depth To End Averages [nm] | | Jump 3 Target Strain Rate ($\dot{P}/P$) [s$^{-1}$] | |
| Lower Mask | | Jump 4 Target Strain Rate ($\dot{P}/P$) [s$^{-1}$] | |
| Upper Mask | | Final Load [mN] | |

Notes: ___________________________________________________
___________________________________________________
___________________________________________________
___________________________________________________

# Appendix-C

The complete JSON schema for the nanoindentation job traveler which enforces consistent data collection. It features error validation, file upload dialogs for associating with metadata, and an embedded code (operation) for execution upon submission, ensuring analysis accuracy and consistency while adding some automation for routine and established tasks.

```json
{
	"description":"Form to track the nanoindentation (iMicro indenter) details for the strain rate jump test (SRJT) method.",
	"definitions":{
		"files":{
			"$id": "#/definitions/files",
			"properties": {
			  "file": {
				"type": "string",
				"format": "url",
				"options": {
				  "upload": {
					"upload_handler": "uploadHandler",
					"auto_upload": true,
					"allow_reupload": true
				  }
				},
				"links": [
				  {
					"href": "https://drive.google.com/file/d/{{self}}/view?usp=share_link",
					"rel": "view"
				  }
				]
			  }
			},
			"type": "object",
			"title": "file",
			"format": "grid"
		}
	},
	"properties":{
		"path":{
			"type":"string",
			"default":"NI",
			"options":{
				"hidden":true
			}
		},
		"operations": {
			"type": "array",
			"items": {
			  "type": "object",
			  "properties": {
				"operationId": {
				  "type": "string"
				},
				"storageType": {
					"type": "string"
				},
```

```json
            "parameters": {
                    "type": "array",
                    "items": {
                            "type" : "string"
                    }
            },
            "result": {
                    "type": "object",
                    "properties": {
                            "schema": {
                                    "type": "string"
                            },
                            "output": {
                                    "type": "string"
                            }
                    }
            }
        }
    },
    "default": [
            {
                    "operationId" : "arn:aws:ecs:us-east-2:132238614819:task-definition/birdshot-ni-srjt:3",
                    "storageType" : "GDrive"
            }
    ],
    "options": {
      "hidden": true
    }
},
"Parent ID":{
        "errorMessage":"Please enter a Sample ID such as AAA01_2209_VAM_A_T01 or leave this field blank for DED samples. At this time, the only allowed Parent/Child samples for Nanoindentation are T01/T04 or no parent/C01",
        "format":"custom",
        "pattern":"^$|([A-Z]{3}[0-9]{2})_(22|23)((0[1-9])|(1[0-2]))_(VAM)_([A-Z]+)_(T01)",
        "propertyOrder":0,
        "type":"string"
},
"Sample ID":{
        "errorMessage":"Please enter a Sample ID such as AAA01_2209_VAM_A_T04 or AAA01_2209_DED_A_C01. At this time, the only allowed Parent/Child samples for Nanoindentation are T01/T04 or no parent/C01",
        "format":"custom",
        "pattern":"([A-Z]{3}[0-9]{2})_(22|23)((0[1-9])|(1[0-2]))_(VAM|DED)_([A-Z]+)_(T04|C01)",
        "propertyOrder":1,
        "type":"string"
},
"Operator":{
        "fieldType":"text",
        "format":"text",
        "propertyOrder":2,
        "type":"string"
},
"Indentation Date":{
        "description":"YYYY-MM-DD",
        "errorMessage":"Please enter a date in the YYYY-MM-DD format",
        "format":"custom",
        "pattern":"^202[23]-((0[1-9])|(1[0-2]))-((0[1-9])|([12][0-9])|(3[0-1]))([ -](([01][0-9])|(2[0-3]))(:[0-5][0-9]){1,2})?$",
```

```json
            "propertyOrder":3,
            "type":"string"
    },
    "Indenter Tip Number":{
            "propertyOrder":4,
            "type":"string"
    },
    "Tip Cal Date":{
            "description":"YYYY-MM-DD",
            "errorMessage":"Please enter a date in the YYYY-MM-DD format",
            "format":"custom",
            "pattern":"^202[23]-((0[1-9])|(1[0-2]))-((0[1-9])|([12][0-9])|(3[0-1]))([ -](([01][0-9])|(2[0-3]))(:[0-5][0-9]){1,2})?$",
            "propertyOrder":5,
            "type":"string"
    },
    "Number of Indents":{
            "propertyOrder":6,
            "type":"number"
    },
    "Device Used":{
            "propertyOrder":7,
            "type":"string",
            "enum": [
                    "KLA iMicro Nanoindenter (iMicro 1)"
            ]
    },
    "Time Spent":{
            "description":"HH:MM",
            "errorMessage":"Please enter a time in the HH:MM format",
            "fieldType":"text",
            "format":"custom",
            "pattern":"^[0-9]+:[0-5][0-9]$",
            "propertyOrder":8,
            "type":"string"
    },
    "Sample Poisson Ratio":{
            "type":"string",
            "propertyOrder":9
    },
    "Machine Settings":{
            "properties":{
                    "Target Load":{
                            "description":"mN",
                            "propertyOrder":0,
                            "type":"number",
                            "default":"1000"
                    },
                    "Target Depth":{
                            "description":"nm",
                            "propertyOrder":1,
                            "type":"number",
                            "default":"3000"
                    },
                    "Target CSM Frequency":{
                            "description":"Hz",
                            "propertyOrder":2,
                            "type":"number",
                            "default":"100"
                    },
                    "Surface Approach Velocity":{
                            "description":"nm/s",
```

```
                                "propertyOrder":3,
                                "type":"number",
                                "default":"100"
                        },
                        "Target Dynamic Displacement":{
                                "description":"nm",
                                "propertyOrder":4,
                                "type":"number",
                                "default":"2"
                        },
                        "Hold Max Load Time":{
                                "description":"s",
                                "propertyOrder":5,
                                "type":"number",
                                "default":"0"
                        },
                        "Surface Approach Distance":{
                                "description":"nm",
                                "propertyOrder":6,
                                "type":"number",
                                "default":"2000"
                        },
                        "Data Acquisition Rate":{
                                "description":"Hz",
                                "propertyOrder":7,
                                "type":"number",
                                "default":"100"
                        },
                        "Depth to Start Averages":{
                                "propertyOrder":8,
                                "type":"number",
                                "description":"nm",
                                "default":"2100"
                        },
                        "Depth to End Averages":{
                                "propertyOrder":9,
                                "type":"number",
                                "description":"nm",
                                "default":"2450"
                        },
                        "Lower Mask":{
                                "propertyOrder":10,
                                "type":"number",
                                "default": 100
                        },
                        "Upper Mask":{
                                "propertyOrder":11,
                                "type":"number",
                                "default": 25
                        }
                },
                "propertyOrder":10,
                "type":"object"
        },
        "Test Parameters":{
                "properties":{
                        "Initial Depth":{
                                "description":"nm",
                                "type":"number",
                                "default":"600",
                                "propertyOrder":0
                        },
```

```
				"Target Depth: Jump 1":{
					"description":"nm",
					"type":"number",
					"default":"1000",
					"propertyOrder":1
				},
				"Target Depth: Jump 2":{
					"description":"nm",
					"type":"number",
					"default":"1500",
					"propertyOrder":2
				},
				"Target Depth: Jump 3":{
					"description":"nm",
					"type":"number",
					"default":"2000",
					"propertyOrder":3
				},
				"Target Depth: Jump 4":{
					"description":"nm",
					"type":"number",
					"default":"2500",
					"propertyOrder":4
				},
				"Final Depth":{
					"description":"nm",
					"type":"number",
					"propertyOrder":5
				},
				"Initial Target Strain Rate":{
					"type":"number",
					"default":"0.2",
					"propertyOrder":6
				},
				"Target Strain Rate: Jump 1":{
					"type":"number",
					"default":"0.02",
					"propertyOrder":7
				},
				"Target Strain Rate: Jump 2":{
					"type":"number",
					"default":"0.2",
					"propertyOrder":8
				},
				"Target Strain Rate: Jump 3":{
					"type":"number",
					"default":"0.002",
					"propertyOrder":9
				},
				"Target Strain Rate: Jump 4":{
					"type":"number",
					"default":"0.2",
					"propertyOrder":10
				},
				"Final Load":{
					"type":"number",
					"propertyOrder":11
				}
			},
			"propertyOrder":11,
			"type":"object"
		},
```

```
            "Raw Data File Upload":{
                    "propertyOrder":12,
                    "title": "Raw Data (.nmd) File Upload",
                    "description": "Please attach the raw data (.nmd) file.",
                    "type": "array",
                    "$ref":"#/definitions/files"
            },
            "Data File Upload":{
                    "propertyOrder":13,
                    "title": "Data (.xlsx) File Upload",
                    "description": "Please attach the data (.xlsx) file to be processed.",
                    "type": "array",
                    "$ref":"#/definitions/files"
            },
            "Notes":{
                    "type":"string",
                    "propertyOrder":14
            }
        },
        "title":"Nanoindentation (SRJT) Details",
        "type":"object"
}
```